# Reliability-based lifetime fatigue damage assessment of offshore composite wind turbine blades

*Chizhi Zhang[1], Hua-Peng Chen [2*], Kong Fah Tee[3] and Dongfang Liang[4]*

ABSTRACT: This paper presents a method for stochastic deterioration modelling and fatigue damage assessment for composite wind turbine blades operating in offshore environments. The fatigue damage of the composite blades is analysed and assessed based on the estimates for the applied loads along the blade span, stress analysis, fatigue crack evolution, and lifetime probability of fatigue failure. The complex stress states of the blade are mainly caused by the aerodynamic loads generated by corrected blade element momentum theory, gravity loads and centrifugal loads. The fatigue of the wind turbine blade is then investigated on the basis of the actual fatigue damage propagation process. The stochastic gamma process is introduced to calculate the probability of fatigue failure of the blade for various critical limits, and these results together with lifecycle cost analysis are employed to determine the optimum maintenance strategy. Finally, a numerical example for an NREL 5 MW wind turbine blade is adopted to demonstrate the applicability of the proposed method. The numerical results show that the proposed approach can provide a reliable tool for estimating stress states, evaluating fatigue damage, analysing lifetime fatigue failure probability and optimising repair time of the composite wind turbine blade.

[1] Institute for Smart Transportation Infrastructure, East China Jiaotong University, Jiangxi 330013, China; Department of Engineering, University of Hull, HU6 7RX, UK; Department of Engineering, University of Cambridge, CB2 1PZ, UK; Email: C.Z.Zhang@hull.ac.uk.
[2] Corresponding author, Institute for Smart Transportation Infrastructure, East China Jiaotong University, Jiangxi 330013, China; Email: hp.chen@ecjtu.ed.cn.
[3] Department of Engineering, University of Greenwich, ME4 4TB, UK; Email: K.F.Tee@gre.ac.uk.
[4] Department of Engineering, University of Cambridge, CB2 1PZ, UK; Email: DL359@cam.ac.uk.

## Introduction

Due to the increasing demand for electrical energy, the researchers are now looking at the field of renewable energy, such as solar energy, wind energy, tidal energy, and formulating relevant policies to promote their development and utilisation. Wind energy is currently one of the most widely developed new energy sources, and the composite blade is one of the critical components of the wind turbine (Toft and Sørensen 2011). In order to ensure the high reliability of the blade under the wind load, strict design requirements are needed for manufacturing the composite blade. Aerodynamic load, gravitational load, inertial load and operational load are main loading conditions, and they are complicated during the service time in harsh environments (Wang et al. 2016). Based on the stress analysis, the performance assessment can be undertaken to investigate the fatigue damage, and the failure probability analysis can be utilised for optimising maintenance strategies. Thus, it is crucial to develop a reliable framework for analysing the fatigue damage and failure probability of the wind turbine blade to optimise the maintenance decision.

Fatigue failure is often investigated by both ultimate state assessment and long-term fatigue analysis during the blade design and service stages (Kensche 2006; Kong et al. 2005). The prediction of fatigue damage propagation needs a proper model adopting the rate of fatigue damage growth with cyclic loading in a deterministic form. The failure mechanism of composite blades needs to be investigated first to understand the fatigue propagation of the blade within the service life and to make the fatigue model more reliable. However, this is difficult due to the complication of the mechanical properties of composite materials, and the limited experiments available on the composite blade fatigue test (Yang et al. 2013). In addition, the fatigue assessment and reliability analysis for blades is comprehensive, including blade model creation, random wind field simulations, stress analysis conditions, and fatigue damage evaluation for complex loadings (Hu et al. 2016).

The aerodynamic loading can be generated by the blade element momentum (BEM) theory (Zhang et al. 2018). However, this method is not well-matched with the real field measurement if the axial factor is more than one-third value. In recent studies, the BEM theory has been corrected as turbulent wake state corrections to provide increasingly accurate results, such as linear correction by Spera (1994) and nonlinear correction by Buhl Jr (2005). Besides the turbulent wake state correction, researchers refined the further correction by the mass flux by introducing the loss factor into the function, and a new equation of loss factors (Shen et al.

2005) is developed to fit with the experimental measurement. In addition to aerodynamic loads, the gravity, centrifugal and operation loads on the blade should be considered in the calculation of the blade loadings, by using distributed loads applied on the deformed blade structure (Wang et al. 2014).

Current research on fatigue damage assessment typically uses load spectra applied on finite element (FE) models of wind turbine blades, which estimates the numerical results of the locations near the blade root. The FE model of the wind turbine blade is adopted to investigate the maximum stress during various wind speeds. The FE model constructed from design data can be updated by using the measured modal parameters such as natural frequencies through the model updating method (Qu et al. 2018, 2019). Some methods for estimating the fatigue cycles are utilised for fatigue analysis, such as rain flow counting, faster Fourier transfer (Epaarachchi and Clausen 2006; Grujicic et al. 2010; Shokrieh and Rafiee 2006). By combining the Goodman diagram with S-N curve, the fatigue damage for each cycle can be calculated (Mandell et al. 2003; Marín et al. 2009). The total fatigue damage in the blade can be predicted by using either linear or nonlinear models, for example, linear Miner model (Kong et al. 2006) and the power model (Sørensen 2009). The Paris model, integrating with stochastic deterioration modelling for determining optimum maintenance strategy during the blade lifetime, is presented in the study (Chen et al. 2017).

The gamma process accumulates monotonically in one direction over time with an independent non-negative gamma distribution increments with the identical scale parameter. Other stochastic methods, such as Markov chains and Brownian movements, could be used for modelling stochastic deterioration, but may not be suitable for continuous and time-dependent stochastic deterioration modelling. Gamma process is suitable for modelling progressive damage, such as wear, fatigue and corrosion. Thus, the gamma process is appropriate for modelling irreversible and gradual fatigue damage by continuous use of a wind turbine blade (van Noortwijk 2009). In previous studies (Chen and Alani 2013; Huang et al. 2016; Zhang and Tee 2018), the gamma process has been considered as an appropriate stochastic approach to simulate the stochastic deterioration process, such as corrosion in concrete bridge and fatigue damage of wind turbine blades. Gamma process model is an appropriate approach for deterioration since it has been proved to be more versatile and increasingly used in optimal maintenance strategies. The gamma process has been used for the numerical simulations of the long-term creep deformations of the prestressed concrete bridges (Strauss et al. 2017). The probability of failure of wind turbine blades can be estimated based on Bayesian updating for the exisiting data. On the basis of the failure probability of the wind turbine blade, a cost-effective model is proposed for the rational and optimal planning of maintenance actions for composite blades.

The fatigue assessment of the composite wind turbine blades has been studied by various scholars. The existing studies, such as by Kensche (2006), Kong et al. (2005) and Martín et al. (2009), focus on the fatigue analysis of wind turbine blade. However, these studies did not involve further research on time-dependent failure probability analysis as well as risk-based optimum maintenance strategy. The effective maintenance and operation of wind turbine blades in harsh environments require a reliable prediction for lifetime fatigue damage progress subjected to repeated wind loads throughout the service life. In this study, a realistic damage model for composite material together with Bayesian updated gamma process modelling is presented for the composite wind turbine blades, and then the lifetime probability of fatigue failure is estimated for optimising maintenance strategy. It is of great importance for blade maintenance to correctly predict the fatigue damage growth and to effectively determine the maintenance strategy. However, the research on the prediction of the fatigue crack development of wind turbine blades and the studies for evaluating the lifetime failure probability are very limited, especially related to the stress states during operation. Therefore, there is a need to develop an approach for estimating the lifetime probability of fatigue failure and evaluating the risk-based optimum maintenance strategy. The proposed method aims to provide an effective tool for predicting the fatigue failure probability of the deteriorating wind turbine blade and for determining the risk and cost balanced maintenance decisions from lifetime failure probability.

A numerical example of the NREL 5MW wind turbine blades is investigated to estimate the fatigue damage and the maintenance decision by the proposed method. The aerodynamic loads for the wind fields are obtained from the corrected BEM method. Gamma parameters are estimated from the increments by two statistical methods and updated by Bayesian theory by using the existing data. Based on the updated failure probability, the expectation of the total cost during operation lifetime can be minimised to optimise the maintenance strategy. The numerical example shows that the gamma process is a reliable tool for modelling the fatigue damage of the wind turbine blade and determining the probability of failure during the service time. From the results, the proposed method can provide a useful tool for evaluating the performance of the wind turbine blade, and assess the fatigue damage assessment performance of the composite wind turbine blades in the harsh environments of offshore wind fields.

## Fatigue damage process of wind turbine blades

Fatigue analysis has been regarded as a critical part of the wind turbine blade design process. In this study, the force of the wind turbine blade is established for different wind speeds, and the results of the FE model are used for the fatigue analysis process. Fatigue life is then estimated by the counting method and the fatigue model. By combining these methods, stochastic models can be used to analyse the lifetime probability of fatigue, which is useful for optimising maintenance decisions.

### Stress loading matrix on the blade

The stress applied to each element should be analysed to assess the fatigue damage in the blade. The loading for importing to the FE model can be used as a loading matrix to simulate the blade performance. Each element of the blade sums the aerodynamic load, gravity load and centrifugal load on the direction of edgewise force $dF_x$ and flapwise force $dF_y$, expressed as

$$\begin{bmatrix} dF_x \\ dF_y \end{bmatrix} = \begin{bmatrix} dF_{ax} + dF_{gx} + dF_{cx} \\ dF_{ay} + dF_{gy} + dF_{cy} \end{bmatrix} \tag{1}$$

The flow chart for the blade performance assessment process is shown in Figure 1.

#### *Aerodynamic model*

The BEM theory is a standard mathematical method to analyse the aerodynamic loading of the wind turbine on the blades by assuming that different control volumes are independent. The one-dimensional momentum method can calculate the wind load on the wind turbine under different incoming wind speeds, wind blade rotation speeds and blade attack angles, so that the axial thrust, torque and the power of the wind turbine can be obtained.

The details of the BEM method is described in detail in the thesis (Zhang 2018). The experimental data by Moriarty and Hansen (2005) show that BEM function does not agree with the experimental data for axial induction factors with a value of more than 0.4. To describe the relationship for higher axial inductor factor, turbulent wake state corrections are discussed as a proper way to solve the large axial inductor value problem. According to existing studies, two turbulent wake states of the wide corrections are presented, the classical linear Spera's correction (Spera 1994) and new nonlinear Buhl's correction (Buhl Jr 2005), respectively. The tip loss corrections (Koh and Ng 2016; Shen et al. 2005) is also introduced into the function because a reduced torque is generated at the tip of the wind turbine blade.

The BEM model can be derived by using an iterative procedure. The results for different corrected models of the thrust coefficient with experimental data of the axial induction factor are shown in Figure 2. From the results, the Spera's correction over-predicts the results, whereas the Buhl's correction under-predicts the results, compared with experimental data. The results also show that the trends of the corrections give reasonable estimates for larger values of the axial induction factor. However, the Spera's correction is slightly different from the measurements for higher loss factors. The reason for this is that the correction of Spera is linear, and the critical induction factor is small.

From the BEM method, the aerodynamic loads can be obtained. After applying the aerodynamic load to the global coordinate, the edgewise and flapwise forces for each blade element are given as

$$\begin{bmatrix} dF_{ax} \\ dF_{ay} \end{bmatrix} = \begin{bmatrix} -sin\varphi & cos\varphi \\ sin\varphi & -cos\varphi \end{bmatrix} \begin{bmatrix} dL_f \\ dD_f \end{bmatrix} \tag{2}$$

where $\varphi$ is the local relative wind angle, and $dL_f$, and $dD_f$ are lift and drag force in each element, respectively.

***Gravity loads in the FE model***

In the structural analysis of wind turbine blades, the gravity should be considered since the mass of the blade is significant and could not be ignored. The gravity load vector of the blade in the global coordinate is given by

$$\begin{bmatrix} dF_{gx} \\ dF_{gy} \end{bmatrix} = \begin{bmatrix} -cos\delta \\ sin\delta \end{bmatrix} [g\mu] \tag{3}$$

where $\delta$ is coned angle; $g$ is the gravity constant; and $\mu$ is the mass per unit length of each blade element.

***Centrifugal loads in the FE model***

Because of the rotation and the mass of blade, centrifugal loads are also important during the service life. The centrifugal load vector of the blade in the global coordinate is given by

$$\begin{bmatrix} dF_{cx} \\ dF_{cy} \end{bmatrix} = \begin{bmatrix} cos\delta \\ sin\delta \end{bmatrix} [r\Omega^2\mu] \tag{4}$$

where $r$ is the rotor radius, and $\Omega$ is the rotation speed.

### Fatigue damage assessment process

Three main loads can be obtained from the above approaches, and by using these loads, the fatigue damage can be analysed based on the force analysis solutions at different speeds. The process for assessing blade fatigue damage is followed by Weibull distribution to simulate wind speed, the number of fatigue cycles, and the Goodman diagram correction of the load spectrum. The *S-N* curve calculates each fatigue damage, which is implemented into the fatigue model for calculations. Figure 3 shows the flow chart of this process.

### Fatigue damage model

Apart from linear and exponential fatigue damage propagation models, other nonlinear damage propagation models are also available. The fatigue damage in composites contains both microscopic and macroscopic mechanisms at all stages during the fatigue process. During the initial period of the lifetime of composite blades, small non-interactive cracks occur in the matrix, and some broken fibres begin to appear. With the matrix cracking density reaching saturation and fibre breaking continuously, some cracks are coupling, and interfacial debonding occurs in the composites. At the later stage, delamination between layers occurs after crack intersecting each other. Delamination and localised fibre breaking develop rapidly, and the material fractures during the end period of fatigue life.

The fatigue damage evolution for a composite blade could follow a nonlinear mathematic power law in the degradation process, written here as

$$d(t) = at^b \propto t^b \tag{5}$$

The trends of fractures stage depend on the exponential parameters. When the damage reaches the fracture stage under cyclic loading, the entire structure cannot withstand any further stresses, leading to destruction of the structure. At this stage, the release of strain energy is enough to allow the damage to propagate until complete damage. This stage begins with the uncontrolled damage exceeding the critical limit, and it can be reflected in the overall expansion of the structure.

## Stochastic deterioration modelling

Gamma process is an appropriate approach for deterioration and estimation of the probability of failure during the lifetime. Fatigue damage is a process under various loads such as wind load. Therefore, the fatigue damage evolution can be reproduced by the gamma process for simulating the deterioration process.

**Nonstationary gamma process**

The stochastic gamma process for fatigue damage development with time can be used for reproducing the damage growth for composite blades. Here, fatigue damage growth is considered as a stochastic process $\{D(t), t\geq 0\}$ with a random quantity of the increments as time dependencies. The probability density function of the gamma process $Ga(d|v,u)$ is given by

$$Ga(d|v,u) = \frac{u^v}{\Gamma(v)} d^{v-1} e^{-ud} \tag{6}$$

where $v$ is the shape parameter, or can be written as $v(t)=ct^b$ over time $t$ where coefficients $c$ and $b$ are to be determined, and $u$ is the scale parameter. The degradation process is $d(t)= at^b =ct^b/u$ where coefficient $a= c/u$. The $\Gamma(v)$ is complete gamma function $\Gamma(v) = \int_0^\infty x^{v-1} e^{-x} dx$.

It is assumed the deterioration with time $t$ ($t\geq 0$) $D(t)$ is a nondecreasing and continuous function for $t\geq 0$, with $D(0)\equiv 0$. The random quantity $\{D(t), t\geq 0\}$ follows three peculiarities (Huang et al. 2016): (1) $D(0)=0$ with probability one; (2) $D(t)$ has independent increments; (3) $D(t)-D(s)\sim Ga(v(t-s),u)$ for all $t>s\geq 0$. Thus, the mean value and variance of $D(t)$ can be expressed as

$$E\big(D(t)\big) = \frac{v}{u}, \quad Var\big(D(t)\big) = \frac{v}{u^2}, \quad CoV(D(t)) = \frac{\sqrt{Var(D(t))}}{E(D(t))} = \frac{1}{\sqrt{v}} \tag{7}$$

**Lifetime reliability analysis**

For the composite blades, the fatigue failure is defined as experiencing $N$ times loading at full-service time $T$ for wind turbines, depending on the design requirement and environmental conditions, where fatigue damage $D$ reaches the critical unit. According to the relationship between resistant stresses and loading cycles, the bearing capacity of the structure decreases when the number of loading cycles increases. Therefore, the fatigue failure probability of the structure also increases, while the resistance of the composite blades deteriorates. Maintenance and repair for the blade should be undertaken in time to prevent structural failure.

The probability of fatigue failure of composite blades during their lifetime is given as

$$F(t) = Pr\{t \geq t_{cr}\} = Pr\{D \geq D_{cr}\} = \int_{D_{cr}}^{\infty} f(D)dD = \frac{\Gamma(v(t),uD_{cr}))}{\Gamma(v(t))} \tag{8}$$

where $D_{cr}$ is the assumed critical damage value at time $t_{cr}$, depending on the maintenance requirements; $v(t)$ is shape function and can be assumed as time-dependent, i.e. $v(t)= u\cdot d(t)$.

The probability of failure at the $j$th time interval $t_j$ is thus computed from

$$p_j = F(t_j) - F(t_{j-1}) \tag{9}$$

If the fatigue damage $D$ reaches the critical value $D_{cr}$, the probability of fatigue failure at this time $t_{cr}$ becomes a unity, and the composite blade fails. As fatigue damage $D$ approaches the critical value, the requirement for maintenance becomes necessary to reduce the possibility of structural failure and to avoid the intolerable possible failure loss. The service time of composite blades can be extended by proper maintenance policy, based on the lifetime of failure probability.

In the gamma process model for the fatigue damage, the shape function $v(t)$ and scale parameter $u$ should be determined from inspection data. Two commonly used methods available for parameter estimation are the maximum likelihood method and the moment method. These two methods for deriving estimated parameters are described by Mahmoodian and Alani (2014).

**Method of maximum likelihood for parameter estimation**

The maximum likelihood method can estimate the shape parameter and scale parameter by maximising the likelihood function of the gamma distribution increments. In the maximum likelihood method, a fixed set of inspected data gives the highest probability for the values of the model parameters under statistics by maximising the logarithmic likelihood function in gamma process. It is assumed that there are observations $x_1, x_2, \ldots, x_i$ and the population with a probability density function of $GP_x(x_1, x_2, \ldots, x_i; v)$. The maximum likelihood method finds the value for unknown parameter where the observed data occurs at most likely happening.

In gamma process, a set of fatigue damage data $x_i$ ($i=1,2\ldots, j$) with inspection times $t_i$, where $0= t_0 < t_1 < t_2 < \cdots < t_j$, are used for parameter estimation. The observed fatigue damage increments in the likelihood function are $\delta_i = x_i - x_{i-1}$ with a set of independent gamma densities

$$l(\delta_1, \ldots, \delta_j | v, u) = \prod_{i=1}^{j} GP_{X(t_i)-X(t_{i-1})}(\delta_i) = \prod_{i=1}^{j} \frac{u^{v(t_i{}^b - t_{i-1}{}^b)}}{\Gamma\left(v(t_i{}^b - t_{i-1}{}^b)\right)} \delta_i{}^{v(t_i{}^b - t_{i-1}{}^b)-1} e^{-u\delta_i} \tag{10}$$

By computing logarithm the likelihood function and then take the first partial derivatives for $v$ and $u$ of the increments, respectively (Edirisinghe et al. 2013; Mahmoodian and Alani 2014), the estimation of parameters $v_e$ and $u_e$ can be obtained by letting the derivative of $v$ and $u$ in maximum likelihood function be zero, giving

$$u_e = \frac{v_e t_j^{\ b}}{x_j} \tag{11a}$$

$$\sum_{i=1}^{j} (t_i^{\ b} - t_{i-1}^{\ b})(\psi(v_e(t_i^{\ b} - t_{i-1}^{\ b})) - \ln \delta_i) = t_j^{\ b} ln \frac{v_e t_j^{\ b}}{x_j} \tag{11b}$$

where the $\psi(x)$ is the derivative of the logarithm of the gamma function, defined as

$$\psi(x) = \frac{\Gamma'(x)}{\Gamma(x)} = \frac{\partial \ ln\, \Gamma(x)}{\partial x} \tag{12}$$

**Method of moments for parameter estimation**

In statistics, the method of moments is a method of estimating overall parameters such as mean and variance, when the set of inspection data is assumed to follow the distribution. By equating the moments of sample data with population moments, the quantities can be estimated by solving those equations. In the gamma process, the expected value and variance of the accumulated fatigue damage at time $t$ are given by

$$E(D(t)) = \frac{vt^b}{u}, \quad Var(D(t)) = \frac{vt^b}{u^2} \tag{13}$$

It is defined the transformed times between inspections as $\Delta d_i = d(t_i) - d(t_{i-1})$, and for mathematical convenience, $\Delta D_i = D_i - D_{i-1}$ for $i=1, ..., j$. The deterioration increment $\Delta D_i$ has a gamma distribution with shape parameter $v(t_i)$ and scale parameter $u$ for all $i=1, ..., j,$ and the increments $\Delta D_1, ..., \Delta D_j$ are independent, where $D_i$ and $\Delta D_i$ are random quantities of increments and $d_i$ and $\Delta d_i$ depend on current observations. According to the study in (van Noortwijk 2009), the method of moments estimates $v_e$ and $u_e$ as

$$\frac{v_e}{u_e} = \frac{\sum_{i=1}^{j} \delta_i}{\sum_{i=1}^{j} w_i} = \frac{x_j}{t_j^{\ b}} = \bar{\delta} \tag{14a}$$

$$\frac{x_j}{u_e}\left(1 - \frac{\Sigma_{i=1}^{j} w_i^{\ 2}}{(\Sigma_{i=1}^{j} w_i)^2}\right) = \sum_{i=1}^{j} \left(\delta_i - \bar{\delta} w_i\right)^2 \tag{14b}$$

It is obvious that the first equation in the maximum likelihood estimation is the same as the first equation in the moment method estimation. By comparing two methods, the method of moments can produce simple equations for the parameter estimation.

**Bayesian updated parameters**

The likelihood function for the observed deterioration increments $\Delta d_i = d - d_{i-1}$ is assumed to be a product of independent gamma densities. Here, it is assumed that $\lambda$ has a prior $\pi(v)$ and $\pi(u)$ follows Gamma distribution. Thus, Bayesian theory can be rewritten as

$$\pi(v, u | \Delta d_1, \ldots, \Delta d_n) = \frac{l(\Delta d_1, \ldots, \Delta d_n | v, u)\pi(v,u)}{\int_0^\infty \int_0^\infty l(\Delta d_1, \ldots, \Delta d_n | v, u)\pi(v,u) dv du} \tag{15}$$

where $l(\Delta d_1,...,\Delta d_j|v,u)$ is the likelihood function of the inspection data $\Delta d_1,...,\Delta d_j$ when the parametric vector $(v,u)$ is given; $\pi(v,u)$ is the prior density of $(v,u)$ before observing the inspection data; $\pi(v,u|\ \Delta d_1,...,\Delta d_j)$ is the posterior density of $(v,u)$ after observing the inspection data; and $\pi(\Delta d_1,...,\Delta d_j)$ is the marginal density of the inspection data.

By combining with maximum likelihood, the updated gamma process parameters of marginal density can be estimated from

$$\begin{cases}\pi(v|\delta_1,...,\delta_n) = \int_0^\infty \pi(v,u|\delta_1,...,\delta_n)du \\ \pi(u|\delta_1,...,\delta_n) = \int_0^\infty \pi(v,u|\delta_1,...,\delta_n)dv\end{cases} \tag{16}$$

By fitting these updated marginal density functions into gamma functions, the Bayesian updated parameters can be obtained.

### Maintenance cost model

Maintenance can be modelled as a discrete-time process to restore the structure to its original state. There are two typical types of maintenance, i.e. preventive maintenance before failure and corrective maintenance after failure. The maintenance cost model is based on minimising the function related to the preventive maintenance cost $C_P$ and corrective maintenance cost $C_F$. The expected cost model is expressed as

$$C_d(k) = \mathrm{C}_P \cdot (1 - F(t)^y) + \mathrm{C}_F \cdot F(t) \tag{17}$$

where $k = 1, 2, 3\ldots$ represents the number of time intervals to be determined; $y$ is the discount rate for the preventive maintenance cost. The optimal maintenance time interval $k^*$ is then obtained by minimising the expected discounted costs over the lifetime (Chen and Alani 2013).

## Numerical example

### The blade geometry data

In order to demonstrate the applicability of the proposed method for the aerodynamic rotor performance assessment and fatigue damage analysis of composite wind turbine blades, the 5 MW reference wind turbine by NREL is chosen as a reference model (Jonkman et al. 2009). The NREL reported this 5 MW reference wind turbine is a conventional three-bladed and 126 meters dimension upwind turbine. The details of this turbine are shown in Table 1. The aerodynamic parameters of 5 MW wind turbine blade, e.g., aerofoil type, chord length, and twist angle, are given in the report (Jonkman et al. 2009; Kim et al. 2013; Resor 2013). In addition, the specifications of the reference model are introduced as the same input parameters as the above 5 MW blade design data to verify the proposed method in this study.

To compare the FE numerical results under the same environmental conditions and different wind speeds, the aerodynamic data of the aerofoil is used in the reference blade (Jonkman et al. 2009). The aerodynamic performance under normal conditions is analysed by the BEM method using these data. In order to keep the stability of the output power, the rotor speed and pitch angle are changeable during operation. The rotor speed and the pitch angle between the cut-in and cut-out wind speeds are based on optimised data in design codes.

The blade geometry model is generated by connecting 17 aerofoils and spars, which smoothers the transition from section to section and reduces the stress concentration, as shown in Figure 4. After the blade geometry model is generated, the 3D blade model can be imported into FE analysis software. The FE mesh is generated by the FE analysis software for the composite blade model, and the layered shell element is chosen for structural analysis of the model.

Since the cut-in wind speed and cut-out wind speed are 3 m / s and 25 m / s, respectively, three types of BEM methods and their corrections are used in NREL 5 MW wind turbine blades to obtain aerodynamic loads. For different wind speeds, the edgewise force, flapwise force and pitching moment along with the overall coordinates of the blade are calculated in each element of the wind turbine blade. In this study, the existing curves of the NREL 5MW wind turbine model by Jonkman et al. (2009) are used as a benchmark.

Figure 5 shows the results for the predicted thrust, torque, maximum bending moment and power curves from different BEM methods and the relevant experimental data. The results show that the traditional BEM method without correction and the methods with different corrections are in general consistent with the actual design thrust, torque, maximum bending moment and power curve of NREL 5MW wind turbine. Thus, it is reasonable to predict the wind turbine performance by the BEM method from the comparison with the measured rotor curves. From the results in Figure 5 the range of wind speed is divided into four stages, i.e. the lower wind speed range (3-7 m/s), the lower-middle wind speed range (7–12 m/s), the middle-higher wind speed range (12 to 21 m/s) and the higher wind speed range (21–25 m/s).

It is also evident that when the wind speed is in the lower wind speed range, the BEM method without correction gives noticeable errors, especially in the thrust curves and maximum bending moment curves. When the wind speed is lower, the axial induction coefficient becomes higher, resulting in a significant difference, since the traditional BEM method without correction has the problem of the high value of the axial induction coefficient. The results of the BEM method using Spera's correction and Buhl's correction are more appropriate at lower wind speeds, since the BEM method may not be suitable for the case where the value of axial induction factor is relatively high. Therefore, the results for low wind speeds produced by the traditional BEM method without correction are generally lower than those from the corrected BEM methods.

By comparing the power curves predicted by the three BEM methods with the corresponding design curves, these predicted power curves show the same trends as the design power curve in the wind speed range of 8 to 22 m/s. For the predicted power curve by the BEM method, in the lower wind speed range, the predicted curve deviates slightly from the design data. In the lower-middle wind speed range, the predicted curve is consistent with the design curve, but slightly over design data. In the middle-higher wind speed range, the predicted power level is greater than the design data. Finally, in the higher wind speed range, the predicted power level is slightly less than the design data.

The power curve predicted by the BEM method with Buhl's correction is closer to the measured data in the normally operating wind speed range (7–25 m/s). Although the corrections almost have no significant effect on the power curve or energy production, it is important to consider the predicted power curve during startup or shutdown operations (3–7 m/s). Overall, the BEM method with Buhl's correction agrees better the experimental data within this range. From the results, the BEM method with Buhl's correction has the highest correlation with the design data for both low wind speed and normal wind speed. Thus, it can be adopted for reliable fatigue analysis of this reference blade.

**FE modelling of the composite blade**

After the BEM methods generate aerodynamic loads, the forces can be applied to the FE model of the blade, and then the maximum stress can be investigated by FE analysis software. The composite wind turbine blade can be modelled as a 3D shell structure and the element type Shell181 in Ansys is adopted for the structural analysis. In this study, the maximum stress is used for the loading cycles to assess the fatigue damage in wind speed field. Figure 6 shows an example of load distributions and the stress contour for the NREL 5 MW wind turbine blade for the case with a wind speed of 11 m/s.

According to the results of the numerical simulations for the wind turbine from cut-in to cut-out wind speed, the maximum stress curves against the wind speed, i.e. the stress-wind speed curves, are obtained, as shown in Figure 7. The fitted curve is obtained simply by linearly linking two adjacent points from the FE results. As the wind speed increases, the maximum stress also increases. However, after the wind speed reaches the design value, the maximum stress becomes stable. When the wind speed exceeds the design wind speed of the wind turbine, the angle of attack decreases, causing the wind surface to decrease. When the wind speed and turbulence are high, the wind turbine maintains a stable angular velocity to produce stable power.

**Fatigue damage estimation**

The Weibull distribution of European offshore wind farms is assumed here throughout the year, where the shape parameter of 2.0 and the scale parameter of 15 are used in the distribution for the reference blade. Therefore, the wind field simulations within the time period of one hour can be randomised at the interval of one minute, as shown in Figure 8. Through wind speed simulations, the time spectrum of the maximum stress in this given period is obtained.

The number of load cycles can be obtained by the rain flow counting method. Each cycle has a maximum stress $\sigma_{max}$ and minimum stress $\sigma_{min}$, so that the average stress $\sigma_m$ and amplitude $\sigma_a$ of each cycle can be calculated. Figure 9 shows the results of rain flow counts within one hour of a given wind field simulation through Weibull distribution. In order to obtain the fatigue stress cycle within this measurement time, the rain flow counting method is adopted to find the force cycles within one hour, namely, 1771 cycles per hour. The value of the fatigue stress cycle can be utilised to analyse the fatigue damage of wind turbine blades.

For the composite blade, the ultimate stress $\sigma_u$ = 100MPa in the Goodman diagram can be estimated, and the equivalent stress $\sigma_e$ can be obtained at $\sigma_{eqm}$=0. According to the *S-N* curve, the number of life cycles $N_i$ at $\sigma_e$ can be calculated. When the ratio is about 0.18, the total number of cycles of wind turbine blades is about 10 million times and the effective stress that causes fatigue damage is 18 MPa from the *S-N* curve. Therefore, after removing the stress cycles that do not cause fatigue damage, which is less than the effective stress, the effective cycle is 53.5 in the given one hour period, providing the probable total service life of 21.3 year without maintenance.

The cyclic fatigue damage is cumulated, and the fatigue damage propagates in the composite blade. When the damage accumulates to a certain critical value, such as 80% of the ultimate value, the blade structure is considered as function failure in operation. According to Wu and Yao (2010), the parameters in Eq. (5) with values of $a$=16.31 and $b$=0.5896 can be fitted in the fatigue model of the composite material.

**Gamma process for estimating fatigue damage**

The experimental data for stochastic fatigue crack growth described in Chen (2018) are utilised for evaluating the parameters of the gamma process by various parameter estimation methods from the monitored data. The estimated fatigue process parameters and scale factor in the gamma process for each method in this study are shown in Table 2.

According to the estimated gamma parameters, the predicted stochastic fatigue damage propagations are presented in Figure 10. The stochastic gamma process predictions show basically consistent with the increase in fatigue damage of the wind turbine blade from the existing data. According to the FE results, the parameters of gamma process can be updated by Bayesian updating approach and reflect the change due to additional data available, which makes the parameters more reliable. Although the gamma process modelling has some differences in the predictions from the deterministic model, the trends of fatigue damage development are very similar. As time increases, the fatigue damage develops by jumping the values of these intervals during the service time. The uncertainties of stochastic fatigue damage evolution prediction curves in Figure 10 can reflect the fatigue damage uncertainties affected by the time and the environments of the composite blade.

**Lifetime probability of failure**

According to the simulation results, the gamma process can be a very reliable tool to model the fatigue damage process. This stochastic model is then adopted to analyse the reliability of composite wind turbine blades. From the inspection data discussed earlier, the Bayesian updating parameters are fitted to the stochastic model. Figure 11 shows the lifetime distribution results of the probabilities of fatigue failure for four different critical fatigue damage limits, namely $D_{cr}$=60%, 70%, 80%, and 90%, respectively. From the results, the shapes of failure probability curves for different critical fatigue damage limits are very similar. The probability of fatigue failure of the composite blade depends on a given critical limit. At any given time, a lower critical limit generates a higher probability of fatigue failure, and vice versa.

Since the design specification requires sufficient service life for offshore wind turbine blades, maintenance strategies during operation are necessary and can effectively increase the service life and reach the design standards.

**Maintenance policy and cost function**

In a preventive replacement strategy, regardless of the condition of the blade, the blade will be replaced if the specified life span is reached. In the cases when maintenance policies are based on the condition of the blade, and the probability of failure, the long-term average costs will be minimised to balance the risk of structural failure. Thus, the proposed average cost per unit time, together with the probability of fatigue failure, can be adopted to evaluate the optimum time for structural repair or replacement.

In this study, the preventive maintenance cost $C_P$ is assumed to be a rate of corrective maintenance cost $C_F$, and $C_F$ is assumed as unity in the cost function, since only relative value is needed in the calculations. The discount rate of 0.7 is assumed and adopted in the cost model for the preventive maintenance cost, which could be adopted for fatigue damage maintenance. Figure 12 (a) illustrates the optimal maintenance time with various cost rates, and Figure 12 (b) presents the optimal maintenance time and total cost when the cost rate is taken as 0.6. From the results, the optimal maintenance time is 18.71 year, which is the optimised repair time to balance the risk of structural failure and the costs for maintenance.

## Conclusions

In this study, the aerodynamic load is estimated by the BEM method and the correction is undertaken by using various methods to match the experimental data available. By correcting blade tip and hub losses as well as the high axial inductor error, the proposed BEM methods are useful for accurately calculating aerodynamic loads of wind turbine blade. Based on the corrected BEM methods, the power between the cut-in and cut-out wind speeds are estimated and then compared with data from other sources. Therefore, the aerodynamic loads obtained from the corrected BEM methods and other main loads can be adapted to reliably analyse the performance of wind turbine blades. This can be further utilised for analysing the finite element model and optimising the maintenance strategies.

The BEM methods with corrections are then utilised in this study to estimate the aerodynamic load, and then consider the gravity load and the centrifugal load at the same time to obtain the forces in the finite element model with different wind speeds. According to the results for the finite element analysis and the random wind field simulations, the fatigue cycle is counted by the rain flow counting method, and the equivalent stress in each cycle is recalculated by the Goodman diagram. The load spectrum and fatigue damage propagation assessment of the wind turbine blade can be obtained by using the proposed method.

From the stochastic gamma process modelling of the fatigue damage of the composite blades during service lifetime, the proposed method provides reliable simulations for the fatigue damage evolution of wind turbine composite blades. The proposed stochastic modelling reasonably evaluates the life distribution of composite blade failure probability, which can be used to assist the inspection and maintenance of composite blades. From the sample spectrum data in the short term, the accumulation of fatigue damage can be obtained during the sample measurement time period, which is used for reliability analysis. Then, on the basis of the probability of fatigue failure of the composite blade, the optimal repair time is determined by minimising the expected cost.

Finally, a numerical example for the NREL 5 MW wind turbine blades is introduced to demonstrate the applicability of the proposed method. From the results, the proposed method can accurately access the fatigue damage evolution of the composite wind turbine blade. Furthermore, the proposed reliability-based method provides reasonable estimates for the probability of fatigue failure of the blade, which can be utilised for determining the optimum maintenance strategy of the composite blade during lifetime operation.

**Data Availability Statements**

Some or all data, models, or code that support the findings of this study are available from the corresponding author upon reasonable request.

**Acknowledgements**

The corresponding author is grateful for the financial supports received from the National Natural Science Foundation of China (Grant No. 51978263) and the Natural Science Key Foundation of Jiangxi Province (Grant No. 20192ACBL20008).

Table 1 Design properties of NREL 5 MW wind turbine blade

| Type | Design data |
| --- | --- |
| Power | 5 MW |
| Blades number | 3 |
| Rotor radius, hub radius and height | 63m, 1.5m, 90m |
| Start, pitch and max wind speed | 3m/s, 11.4m/s, 25m/s |
| Rotor speed | 6.9 rpm to 12.1 rpm |
| Mass | 17740 kg |
| Rotor orientation | Upwind |
| Control | Variable speed, Collective pitch |
| Wind regime | IEC 61400-3 (offshore) |

Table 2 Gamma process parameters estimated from experimental data available

| | **Fitted values** | **Moments method** | **Maximum likelihood method** | **Bayesian updating** |
|---|---|---|---|---|
| *a* (*c*/*u*) | 16.31 | 16.82 | 17.02 | 18.33 |
| *b* | 0.5896 | 0.5896 | 0.5896 | 0.5527 |
| *c* | - | 7.7676 | 6.7696 | 0.2857 |
| *u* | - | 0.4620 | 0.3978 | 0.0156 |